\documentclass[10.5pt]{article}
\usepackage{array, amsmath, amssymb}
\usepackage{multicol}
\begin{document}
\begin{center}
{\Large\bf Epidemic spreading with immunization rate on complex networks}\\
\vspace{0.5cm}
{\large\bf Shinji Tanimoto}\\
{\texttt{(tanimoto@cc.u-kochi.ac.jp)}\\
{Department of Mathematics, University of Kochi,
Kochi 780-8515, Japan.}} \\
\end{center}
\begin{abstract}
We investigate the spread of diseases, computer viruses or information on complex networks and also
immunization strategies to prevent or control the spread. 
When an entire population cannot be immunized and the effect of immunization is not perfect, 
we need the targeted immunization with immunization rate. 
Under such a circumstance we calculate epidemic thresholds for the SIR and SIS epidemic models. 
It is shown that, in scale-free networks, the targeted immunization is effective only if
the immunization rate is equal to one. We analyze here epidemic spreading on directed
complex networks, but similar results are also valid for undirected ones.
\end{abstract}
\vspace{0.3cm}
\begin{multicols}{2}
\begin{center}
{\bf\large 1. Introduction}  \\
\end{center}
\indent
\indent
The spread of diseases, computer viruses or information
in complex networks such as social networks, the WWW and Internet 
has attracted researchers' attention in various fields. We refer to [2, 3, 8] and
references therein for abundant examples of complex networks.
Also preventing the spread has extensively been investigated.
A typical method to prevent or control an epidemic spread is immunization such as 
vaccination [5]. Generally,
it is difficult or impossible to perfectly immunize an entire population, so
the object or target of immunization must be prescribed.
However in turn, all of target individuals may not be immunized due to, for example, lack of plenty of time and 
resources, otherwise because some are overlooked. We describe this situation as
"imperfect immunization to targeted nodes". Nodes are sometimes called vertices or individuals.
In another aspect, vaccination or a treatment cannot
provide sufficient immunity upon target nodes, which 
we call "insufficient immunity". Therefore, we have to introduce a parameter $\alpha$ ($0 \le \alpha \le 1$)
called immunization rate for such both cases. 
As in [12] we assume that $\alpha$ is a constant throughout. \\
\indent
In [10] some immunization schemes are evaluated in the SIS model
on undirected networks and [4] also offers other schemes for immunization, and [12] provides
targeted immunization strategies in the SIR model. All those studies are devoted to 
undirected. 
In undirected networks links (or edges) do not have directions or they are bidirectional by nature. \\
\indent
However, one can find many directed complex networks [2, 3, 8] in nature, society, 
and artificial structures; food webs, phone-call networks, the WWW, {\it etc}.
The indegree of a node is the number of incoming links into the node and the outdegree is that of
outgoing links emanating from it. If the indegree and outdegree distributions of a directed network follow
power laws: 
\[
P(k) \propto k^{-\gamma}~~ {\rm and} ~~Q(\ell) \propto {\ell}^{-\gamma'}
\]
for indegrees $k$ and outdegrees $\ell$, and if
the exponents satisfy $2 < \gamma, \gamma' \le 3$, then the network is called scale-free.
Undirected networks make no distinction between both degree distributions. \\
\indent
The direction plays an important role in the study of epidemic spreading on a directed network.
Diseases, computer viruses or information are transmitted to other nodes (or vertices) through
outgoing links and a node is infected by incoming links. We derive the critical infection rate or 
the threshold, above which a disease spreads in the networks and below which it dies out. \\
\indent
First we prescribe a target set of immunization in terms of indegrees and outdegrees
on directed networks.
We derive the thresholds for the SIR and SIS epidemic dynamics involving the target set and
an immunization rate $\alpha$. Actually they turn out the same for both. 
In the SIR model, the averaged fraction of nodes that are ever infected
until the disease dies out is also given. 
Finally, based on thresholds, we will discuss whether immunization strategies are effective or not
in scale-free networks. In particular, all targeted immunization schemes are shown to be effective only if
$\alpha = 1$, which was assumed in [10] from the beginning. \\
\indent
Since undirected networks are simpler than directed ones, we immediately see that
similar results hold on undirected networks by the arguments 
presented here, and they are also partially found in [4] for the SIS model and in [12] for the SIR model.\\
\begin{center}
{\bf\large 2. Models of immunization in complex networks}
\end{center}
\indent
\indent
In this section we introduce preliminary definitions and model epidemic spreading with 
immunization rate. Our model is based on [11],
where epidemic spreading with no immunization is discussed on directed networks. 
First we treat the SIR model and later proceed to the SIS model which is regarded as a special case
of the SIR model. \\
\indent
In the SIR model, nodes of the network are divided into the following three 
groups regarding infection states ([5], [7, Chap. 10]): 
Susceptible (S), Infected (I) and Removed (R). In order to deal with immunization such as 
vaccination, we add one more state: Immunized. Outgoing links emanating from immunized nodes 
are harmless.\\
\indent
Hereafter we will denote a susceptible node by
an S-node {\it etc.}, for short.
An S-node becomes infected at a rate of $\lambda$ ($0 \le \lambda \le 1$). 
The parameter $\lambda$ is the infection rate, for which we will derive the critical value 
for an outbreak of a disease, a computer virus, {\it etc.}
The disease can be passed from I-nodes to S-nodes following only the direction of directed links. 
R-nodes have either recovered from the disease or died and so they cannot pass the disease to others.
An I-node becomes an R-node at a rate $\delta$ ($0 \le \delta \le 1$) and, without loss of generality, 
we will set $\delta = 1$. \\
\indent
Let $p(k, \ell)$ be the joint probability distribution of nodes with indegree $k$ and outdegree $\ell$,
and let us denote the marginal distributions by
\[
P(k) = \sum_{\ell} p(k, \ell), ~~ Q(\ell) = \sum_{k} p(k, \ell)
\]
and the averages by
\begin{eqnarray*} 
\langle k\rangle \! \! \! \!& = & \! \!\! \!\sum_{k, \ell} k p(k, \ell) = \sum_{k} k P(k),  \\
\langle{\ell}\rangle \! \! \! \!& = & \! \!\! \! \sum_{k, \ell} {\ell} p(k, \ell) = \sum_{\ell} {\ell} Q(\ell), \\
\langle{k\ell}\rangle \! \! \! \!& = & \! \!\! \! \sum_{k, \ell} k\ell p(k, \ell).
\end{eqnarray*} 
\indent
We denote by $T$ the object or target set for immunization that is characterized in terms of 
indegrees and outdegrees, $\bar{T}$ being the complement of $T$ that is not an object of immunization. 
The notation $(k, \ell) \in T$ means that 
the population of all nodes with indegree $k$ and outdegree $\ell$ is an object of immunization.
In case of $T =\{(k, \ell) | k > K\}$, for example,
the population of all nodes with indegrees exceeding a number $K$ is collectively an 
object of immunization.
The summation restricted to all $(k, \ell)$ in $T$ will be denoted by $\sum_T$, and similarly for 
$\bar{T}$ by $\sum_{\bar{T}}$.
Furthermore, the average of the product $k \ell$ over $T$ is denoted by ${\langle k \ell \rangle}_T$:
\[
{\langle k \ell \rangle}_T = \sum_{(k, \ell) \in T} k \ell p(k, \ell)
= \sum_T k \ell p(k, \ell).
\]
\indent
In this paper we assume that the immunization rate $\alpha$ ($0 \le \alpha \le 1$) is a constant as in [12], 
although it can be dependent on $k$ and $\ell$ as well. 
In the following two models A and B, remark that $\alpha$ has different meanings,
althought the same notation is used. 
Within the population of all nodes with indegree $k$ and outdegree $\ell$, variables
$S_{k,\ell}(t), \rho_{k,\ell}(t), R_{k,\ell}(t)$ indicate
the densities of S-, I-, R-nodes at time $t$, respectively. Naturally, by setting $k = \ell$, 
the same arguments are applied to undirected networks. 
\begin{center}
{\bf A. Imperfect immunization to targeted nodes}
\end{center}
\indent
\indent
This is the case where all of target nodes in $T$ cannot be immunized, because 
some are overlooked or hidden {\it etc.}, and $\alpha$ is like the vaccine coverage within $T$. \\
\indent
Since there are no rewiring of links and immunized nodes are no longer S-nodes, we have 
\begin{eqnarray}
S_{k,\ell}(t) + \rho_{k,\ell}(t) + R_{k,\ell}(t) = \left\{\begin{array}{ll}
1-\alpha, & (k, \ell) \in T,\\
1, & (k, \ell) \in \bar{T},
\end{array} \right.
\end{eqnarray}
The condition $\alpha=1$ implies the perfect immunization for $T$, 
while $\alpha=0$ means no immunization, which coincides with the simple SIR model [11].\\
\indent
Following the dynamical mean-field theory ([6, 9, 10]), we see that
the spreading process on a directed network can be described by the system of differential equations:
\begin{eqnarray} 
\frac{dS_{k,\ell}}{dt} \! \! \! \!& = & \! \!\! \! - \lambda k S_{k,\ell}(t)\theta(t), \\
\frac{d{\rho}_{k,\ell}}{dt} \! \!\! \! & =  &\! \! \! \! \lambda k S_{k,\ell}(t)\theta(t) - {\rho}_{k,\ell}(t), \\
\frac{dR_{k,\ell}}{dt} \! \!\! \!  & = &\! \!\! \!  {\rho}_{k,\ell}(t).
\end{eqnarray} 
The term $\lambda k S_{k,\ell}(t)\theta(t)$
in (2) and (3) indicates the fraction of newly infected nodes through $k$ incoming links, while $\theta(t)$
is the probability of contact with I-nodes from whose outgoing links the disease spreads. Hence
$\theta(t)$ can be written in parts as
\begin{eqnarray} 
\theta(t) = \frac{1}{\displaystyle \sum_{k, \ell}\ell p(k, {\ell})} 
\Big(\displaystyle \sum_{T}\ell  
p(k, \ell){\rho}_{k, \ell}(t) \nonumber \\
+ \displaystyle \sum_{\bar{T}}\ell  
p(k, \ell){\rho}_{k, \ell}(t)\Big),
\end{eqnarray} 
for the later use.
Here the denominator is the average $\langle \ell \rangle$ and each directed link is counted twice as one 
outdegree of some node and as one indegree of another. Hence the average outdegree is equal to
the average indegree: ${\langle{\ell}\rangle} = {\langle{k}\rangle}$.
From (1) the initial conditions imposed upon $S_{k,\ell}$ are 
\begin{eqnarray} 
S_{k,\ell}(0) = \left\{\begin{array}{ll}
1-\alpha, & ~(k, \ell) \in T,\\
1, & ~ (k, \ell) \in \bar{T}.
\end{array} \right.
\end{eqnarray} 
The initial values of other variables are assumed to be zero or nearly zero. 
\indent
\begin{center}
{\bf B. Insufficient immunity}
\end{center}
\indent
\indent
This model is applied to the case, for example, where vaccination or a treatment cannot provide
sufficient immunity upon the target nodes in $T$, so the rate
$\alpha~(0 \le \alpha \le 1)$ represents the immunity effect.
The higher it is, the less infected S-nodes in $T$ become.
The differential equations are divided into two parts. First, for $(k, \ell) \in T$, it follows that
\begin{eqnarray} 
\frac{dS_{k,\ell}}{dt} \! \! \! \!& = & \! \!\! \! - \lambda k (1-\alpha)S_{k,\ell}(t)\theta(t), \\
\frac{d{\rho}_{k,\ell}}{dt} \! \!\! \! & =  &\! \! \! \! \lambda k (1-\alpha)S_{k,\ell}(t)\theta(t) - {\rho}_{k,\ell}(t), \\
\frac{dR_{k,\ell}}{dt} \! \!\! \!  & = &\! \!\! \!  {\rho}_{k,\ell}(t).
\end{eqnarray} 
Secondly, for $(k, \ell) \in \bar{T}$, Eqs. (2)--(4) remain valid. 
The probability $\theta$ is the same as (5) and the initial condition imposed upon $S_{k,\ell}$ is 
\begin{eqnarray*} 
S_{k,\ell}(0) = 1
\end{eqnarray*} 
for all $k$ and $\ell$. The initial values of other variables are zero or nearly zero. \\
\indent
The next section analyzes models A and B in directed complex networks with target set $T$ and
with immunization rate $\alpha$. In Section 4 the corresponding SIS models
are considered. Actually all thresholds turn out the same.
It is easy to see that the following arguments can be applied to models A and B considered in
undirected complex networks. \\
\begin{center}
{\bf\large 3. Epidemic spreading in the SIR model}
\end{center}
\indent
\begin{center}
{\bf A. Imperfect immunization to targeted nodes}
\end{center}
\indent
\indent
To begin with we solve Eq. (2) under the initial conditions (6):
\begin{eqnarray}
S_{k,\ell}(t) = \left\{\begin{array}{ll}
(1-\alpha)e^{-\lambda k \phi(t)} & \textrm{for}~(k, \ell) \in T,\\
e^{-\lambda k \phi(t)} & \textrm{for}~ (k, \ell) \in \bar{T},
\end{array} \right.
\end{eqnarray}
where 
\begin{eqnarray*}
\phi(t) = \int_0^t \theta(t')dt'.
\end{eqnarray*}
By (4), (5) and $R_{k, \ell}(0) = 0$ for all $k$ and $\ell$, $\phi(t)$ has an expression: 
\begin{eqnarray}
\phi(t)  \! \! \! \!& = &\! \! \! \! \frac{1}{\langle \ell \rangle}
\big\{\displaystyle \sum_T\ell p(k, \ell)\int_0^t{\rho}_{k, \ell}(t')dt'   \nonumber\\
&&+ \displaystyle \sum_{{\bar{T}}}\ell p(k, \ell)\int_0^t{\rho}_{k, \ell}(t')dt' \big\} \\
 \! \! \! \!& = &\! \! \! \!  \frac{1}{\langle \ell \rangle}
\big\{\displaystyle \sum_T \ell  
p(k, \ell)R_{k, \ell}(t) \nonumber\\
&&+\displaystyle \sum_{{\bar{T}}}\ell  
p(k, \ell)R_{k, \ell}(t) \big\}. \nonumber
\end{eqnarray} 
To derive the differential equation for $\phi(t)$, we use (1), (10) and (11). Then it follows that
\begin{eqnarray}
\frac{d\phi(t)}{dt} \! \! \! \!&= &\! \! \! \! \frac{1}{\langle \ell \rangle}
\big\{\displaystyle \sum_T\ell p(k, \ell)\rho_{k, \ell}(t) 
+ \displaystyle \sum_{{\bar{T}}}\ell p(k, \ell)\rho_{k, \ell}(t)\big\} \nonumber \\
 \! \! \! \!& = &\! \! \! \!  \frac{1}{\langle \ell \rangle}
\big\{\displaystyle \sum_T\ell p(k, \ell)\big(1-\alpha-R_{k, \ell}(t)-S_{k,\ell}(t)\big) \nonumber\\
&&+ \displaystyle \sum_{{\bar{T}}}\ell  
p(k, \ell)\big(1-R_{k, \ell}(t)-S_{k,\ell}(t)\big) \big\} \nonumber \\
 \! \! \! \!& = &\! \! \! \!  \frac{1}{\langle \ell \rangle}
\big\{\displaystyle \sum_T \ell p(k, \ell)\big(1-\alpha-S_{k,\ell}(t)\big)\nonumber\\
&& +\displaystyle \sum_{{\bar{T}}}\ell  
p(k, \ell)\big(1-S_{k,\ell}(t)\big)\big\} -\phi(t) \\
 \! \! \! \!& = &\! \! \! \!  \frac{1}{\langle \ell \rangle}
\big\{\displaystyle \sum_T \ell p(k, \ell)\big(1-\alpha-(1-\alpha)e^{-\lambda k \phi(t)}\big)\nonumber\\
&&+ \displaystyle \sum_{{\bar{T}}}\ell p(k, \ell)\big(1-e^{-\lambda k \phi(t)}\big)\big\} 
-\phi(t)\nonumber \\
\! \! \! \!& = &\! \! \! \!  \frac{1}{\langle \ell \rangle}
\big\{(1-\alpha)\displaystyle \sum_T\ell p(k, \ell)\big(1-e^{-\lambda k \phi(t)}\big)\nonumber\\
&&+ \displaystyle \sum_{{\bar{T}}}\ell p(k, \ell)\big(1-e^{-\lambda k \phi(t)}\big)\big\} 
-\phi(t). \nonumber
\end{eqnarray}
\indent
We are concerned with a steady state of the epidemic spreading, for which
one has a limit 
\begin{eqnarray*}
\Phi = \lim_{t \rightarrow \infty}\phi(t),
\end{eqnarray*}
together with the condition
\begin{eqnarray*}
 \lim_{t \rightarrow \infty} \frac{d\phi(t)}{dt} = 0.
\end{eqnarray*}
Substituting these two into Eq. (12), we get the equation for $\Phi$ as follows:
\begin{eqnarray}
\Phi  \! \! \! \!& = &\! \! \! \!  \frac{1}{\langle \ell \rangle}
\big\{(1-\alpha)\displaystyle \sum_T \ell p(k, \ell)(1-e^{-\lambda k \Phi}) \nonumber\\
&&{} + \displaystyle \sum_{{\bar{T}}}\ell p(k, \ell)(1-e^{-\lambda k \Phi})\big\}.
\end{eqnarray}
\indent
An epidemic outbreak implies that this equation has a solution $\Phi >0$ other than $\Phi = 0$. 
Since the right hand side of (13) is a monotone increasing and concave function of $\Phi$ 
and its value at $\Phi = 1$ is less than 1, the condition for it is
\begin{eqnarray*}
\frac{d}{d \Phi}\frac{1}{\langle \ell \rangle}\big\{(1-\alpha)
\displaystyle \sum_{T}\ell p(k, \ell)(1-e^{-\lambda k \Phi}) \\
+ \displaystyle \sum_{\bar{T}}\ell 
p(k, \ell)(1-e^{-\lambda k \Phi})\big\}{\big|}_{\Phi = 0} \ge 1.
\end{eqnarray*}
Thus the critical threshold $\lambda_{\rm c}$ is expressed as
\begin{eqnarray}
\lambda_{\rm c} =\frac{{\langle \ell \rangle}}
{(1-\alpha)\sum_T k\ell p(k, \ell)
+ \sum_{{\bar{T}}}k\ell p(k, \ell)}.
\end{eqnarray}
Alternatively, using the identity
\begin{eqnarray*}
 \sum_{{\bar{T}}}k\ell p(k, \ell)
= \langle k{\ell}\rangle - \sum_T k\ell p(k, \ell),
\end{eqnarray*}
(14) is equal to
\begin{eqnarray}
\lambda_{\rm c} = \frac{\langle \ell \rangle}
{\langle k{\ell}\rangle - \alpha {\langle k{\ell}\rangle}_T}.
\end{eqnarray}
\indent
In [7, Chap.10] the total number of infected individuals is discussed for the classical SIR model,
which represents the final outbreak size ${\mathcal{O}}$.
In our setting it is the averaged fraction of nodes ever infected
until the disease dies out. Using the solutions (10), this is written as
\begin{eqnarray}
{\mathcal{O}} \! \! \! \!& = &\! \! \! \!  \sum_{k, \ell} p(k, \ell)(1- S_{k, \ell}(\infty)) \nonumber\\
\! \! \! \!& = &\! \! \! \! 1 - \sum_{{T}}p(k, \ell)S_{k, \ell}(\infty)
- \sum_{{\bar{T}}}p(k, \ell)S_{k, \ell}(\infty) \nonumber \\
\! \! \! \!& = &\! \! \! \! 
 1-(1-\alpha)\sum_{T} p(k, \ell)e^{-\lambda k \Phi}
 -\sum_{\bar{T}} p(k, \ell)e^{-\lambda k \Phi}\nonumber \\
\! \! \! \!& = &\! \! \! \!  1- \sum_{k, \ell} p(k, \ell)e^{-\lambda k \Phi}
+ \alpha \sum_{T} p(k, \ell)e^{-\lambda k \Phi}  \nonumber\\
\! \! \! \!& = &\! \! \! \! 
 1- \sum_{k} P(k)e^{-\lambda k \Phi}
 +\alpha\sum_{T} p(k, \ell)e^{-\lambda k \Phi}. 
\end{eqnarray}
The last term of the right hand side provides the effect of an immunization.\\
\indent
As an illustrative example let us consider an immunization of nodes with large indegrees. This means 
$T =\{(k, \ell) | k > K\}$ or
\begin{eqnarray*}
S_{k,\ell}(t) + \rho_{k,\ell}(t) + R_{k,\ell}(t) = \left\{\begin{array}{ll}
1-\alpha, & k > K,\\
1, & k \le K,
\end{array} \right.
\end{eqnarray*}
where $K$ is a certain large value of the indegree. 
The corresponding initial conditions imposed upon $S_{k,\ell}$ are 
\begin{eqnarray*} 
S_{k,\ell}(0) = \left\{\begin{array}{ll}
1-\alpha, & k > K,\\
1, & k \le K.
\end{array} \right.
\end{eqnarray*}
The initial values of other variables are zero or nealy zero. \\
\indent
Using the solution $\Phi$ of Eq. (13) for this case, we obtain the fraction (16) of the outbreak size:
\begin{eqnarray*}
{\mathcal{O}} =
 1- \sum_{k} P(k)e^{-\lambda k \Phi}
 + \alpha\sum_{k > K} P(k)e^{-\lambda k \Phi},
\end{eqnarray*}
by means of the indegree distribution $P(k)$. 
If the indegree distribution follows a power law 
\[
P(k) \propto k^{-\gamma}, ~~k \ge m
\] 
with $\gamma > 1$, then $P(k) = 
(\gamma-1)m^{\gamma-1}k^{-\gamma}~(k \ge m)$,
where $m$ is the minimum indegree.
Defining the incomplete gamma function [1, p.260] by
\[
\Gamma(a,x)= \int_x^{\infty}t^{a-1}e^{-t}dt,
\]
we can write the fraction (16) of the outbreak size as
\begin{eqnarray*}
{\mathcal{O}}\! \!& = & \! \! 1 - \int_m^{\infty}P(k)e^{-\lambda k \Phi}dk 
 + \alpha\int_{K}^{\infty} P(k)e^{-\lambda k \Phi}dk  \\
\! \!& = & \! \! 1- (\gamma -1)(\lambda m \Phi)^{\gamma - 1}
\Gamma(1-\gamma, \lambda m \Phi){} \\
&&{}+ \alpha(\gamma -1)(\lambda K \Phi)^{\gamma - 1}
\Gamma(1-\gamma, \lambda K \Phi),
\end{eqnarray*}
by the continuous approximation.  
\begin{center}
{\bf B. Insufficient immunity}
\end{center}
\indent
\indent
Since similar arguments used in A are also applied to this model, we only give an outline of the analysis.
Solving (7)--(9) under the initial condition $S_{k,\ell}(0) = 1$ for all $(k, \ell)$, 
we have, in place of (10) and (13), 
\begin{eqnarray*}
S_{k,\ell}(t) = \left\{\begin{array}{ll}
e^{-\lambda (1-\alpha)k \phi(t)} & \textrm{for}~(k, \ell) \in T,\\
e^{-\lambda k \phi(t)} & \textrm{for}~ (k, \ell) \in \bar{T},
\end{array} \right.
\end{eqnarray*}
and
\begin{eqnarray*}
\Phi  \! \! \! \!& = &\! \! \! \!  \frac{1}{\langle \ell \rangle}
\big\{\displaystyle \sum_T \ell p(k, \ell)(1-e^{-\lambda (1-\alpha) k \Phi}) \\
&&{} + \displaystyle \sum_{{\bar{T}}}\ell p(k, \ell)(1-e^{-\lambda k \Phi})\big\}, \nonumber
\end{eqnarray*}
respectively. Solving the following equation for $\lambda$,
\begin{eqnarray*}
\frac{d}{d \Phi}\frac{1}{\langle \ell \rangle}\big\{
\displaystyle \sum_{T}\ell p(k, \ell)(1-e^{-\lambda (1-\alpha) k \Phi}) \\
+ \displaystyle \sum_{\bar{T}}\ell 
p(k, \ell)(1-e^{-\lambda k \Phi})\big\}{\big|}_{\Phi = 0}=1,
\end{eqnarray*}
we again obtain (14) and (15). When $T =\{(k, \ell) | k > K\}$ as above,
the averaged fraction of nodes ever infected until the disease dies out becomes 
\begin{eqnarray*}
{\mathcal{O}} =
 1- \sum_{k > K} P(k)e^{-\lambda (1- \alpha)k \Phi} -\sum_{k \le K} P(k)e^{-\lambda k \Phi},
\end{eqnarray*}
by an application of (16). \\ 
\begin{center}
{\bf\large 4. Epidemic spreading in the SIS model}
\end{center}
\indent
\indent
In the SIS model, R-nodes are absent and nodes that recovered from the disease cannot acquire
eternal immunity and may be infected again and again. 
Let $T$ and $\alpha$ $(0 \le \alpha \le 1)$ be the same as before in the models A and B. 
The differential equations for the densities $S_{k,\ell}(t), \rho_{k,\ell}(t)$ of S-, I-nodes at time $t$ are easily
derived from those for the SIR models in Section 2. 
\begin{center}
{\bf A. Imperfect immunization to targeted nodes}
\end{center}
\indent
\indent
In this case, remarking (1), we have
\begin{eqnarray*}
S_{k,\ell}(t) = \left\{\begin{array}{ll}
1-\alpha- \rho_{k,\ell}(t), & \textrm{for}~(k, \ell) \in T,\\
1- \rho_{k,\ell}(t), & \textrm{for}~ (k, \ell) \in \bar{T}.
\end{array} \right.
\end{eqnarray*}
Therefore, Eqs. (2)--(4) in Section 2 are replaced by the differential equations
\begin{eqnarray} 
\frac{d\rho_{k,\ell}(t)}{dt} = \left\{\begin{array}{ll}
\lambda k (1-\alpha-\rho_{k, \ell}(t))\theta(t) - \rho_{k, \ell}(t),  \\
~~~~~~~~~~~~~~~~~~~~~~~~~~~~~~\textrm{if}~ (k, \ell) \in T,\\
\\
\lambda k (1-\rho_{k, \ell}(t))\theta(t) - \rho_{k, \ell}(t),  \\
~~~~~~~~~~~~~~~~~~~~~~~~~~~~~~\textrm{if}~ (k, \ell) \in \bar{T},
\end{array} \right.
\end{eqnarray} 
where $\theta(t)$ is the same probability as (5). \\
\indent
At the steady state, as in Section 3, we will have the condition
\[
\lim_{t \rightarrow \infty}\frac{d{\rho}_{k,\ell}}{dt}  = 0
\]
for all $k$ and $\ell$, and a limit
\[
\Theta = \lim_{t \rightarrow \infty}\theta(t).
\]
So we get from (17),
\begin{eqnarray*} 
\lim_{t \rightarrow \infty}\rho_{k, \ell}(t) = \left\{\begin{array}{ll}
(1-\alpha)\lambda k\Theta/(1+\lambda k \Theta), & \textrm{if}~ (k, \ell) \in T,\\
\lambda k\Theta/(1+\lambda k \Theta), & \textrm{if}~ (k, \ell) \in \bar{T}.
\end{array} \right.
\end{eqnarray*} 
\indent
Substituting these into (5) and letting $t \to \infty$, we have the equation for $\Theta$ as follows:
\begin{eqnarray*} 
\Theta = \frac{1}{\langle \ell \rangle(1+\lambda k \Theta)}
\Big(\displaystyle \sum_{T}(1-\alpha)\lambda k \ell p(k, \ell)\Theta \\
+ \displaystyle \sum_{\bar{T}}\lambda k \ell p(k, \ell)\Theta \Big). 
\end{eqnarray*} 
If this has a solution $\Theta > 0$ other than $\Theta = 0$, then it corresponds to an endemic state.
Since the right hand side of the equation is a monotone increasing
and concave function of $\Theta$ and its value at $\Theta = 1$ is less than 1,
the condition for an endemic outbreak is
\begin{eqnarray*} 
\lefteqn{\frac{d}{d \Theta}\frac{1}{\langle \ell \rangle}
\Big(\displaystyle \sum_{T}\frac{(1-\alpha)\lambda k \ell  
p(k, \ell)\Theta}{1+\lambda k \Theta}}\\
&&{}+ \displaystyle \sum_{\bar{T}}\frac{\lambda k \ell  
p(k, \ell)\Theta}{1+\lambda k \Theta }\Big){\Big|}_{\Theta = 0} \ge 1.
\end{eqnarray*} 
Again, this yields the same threshold (14) or (15) as the model A for the SIR model:
\begin{eqnarray*}
\lambda_{\rm c} = \frac{{\langle \ell \rangle}}
{\langle k{\ell}\rangle - \alpha \sum_T k\ell p(k, \ell)}
= \frac{{\langle \ell \rangle}}
{\langle k{\ell}\rangle - \alpha {\langle k{\ell}\rangle}_T}.
\end{eqnarray*}
\begin{center}
{\bf B. Insufficient immunity}
\end{center}
\indent
\indent
This model just coincides with Eq. (10) of [4] for the SIS model on undirected networks. 
Taking Eq. (3) for $\bar{T}$ and Eq. (8) for $T$, we get
\begin{eqnarray*} 
\frac{d\rho_{k,\ell}(t)}{dt} = \left\{\begin{array}{ll}
\lambda k (1-\alpha)(1-\rho_{k, \ell}(t))\theta(t) - \rho_{k, \ell}(t), & \\
~~~~~~~~~~~~~~~~~~~~~~~~~~~~~~\textrm{if}~ (k, \ell) \in T,\\
\\
\lambda k (1-\rho_{k, \ell}(t))\theta(t) - \rho_{k, \ell}(t), & \\
~~~~~~~~~~~~~~~~~~~~~~~~~~~~~~\textrm{if}~ (k, \ell) \in \bar{T}.
\end{array} \right.
\end{eqnarray*} 
Using this equation for (17) and repeating the above calculations,
we obtain again the same threshold (14) or (15) as in A.
\\
\begin{center}
{\large \bf 5. Immunization strategies}  \\
\end{center}
\indent
\indent
We have seen that the thresholds are all the same in four cases. So in any one of them 
we can evaluate immunization strategies based on the thresholds. In view of (14) or (15),
to control epidemic outbreaks, it is desirable to choose 
the set $T$ so that ${\langle k{\ell}\rangle}_{T}$ becomes as large as possible, and
$T$ should be a considerably small region
compared to the entire population from a practical point of view. \\
\indent
First remark that in case of no immunization ($\alpha = 0$) the epidemic threshold is equal to
\[
\lambda_{\rm c}^{\circ}=\frac{{\langle \ell \rangle}}{\langle k{\ell}\rangle},
\]
by (14) or (15). \\
\indent
If the entire population is an object for immunization, 
that is, $\bar{T}$ is empty, then from the identity
${\langle k\ell \rangle}_{T} = {\langle k\ell \rangle}$
the threshold due to this global immunization is
\begin{eqnarray*}
\hat{\lambda}_{\rm c} =\frac{{\langle \ell \rangle}} 
{{\langle k\ell \rangle} -\alpha{\langle k\ell \rangle}_T}
= \frac{1}{1-\alpha}\frac{{\langle \ell \rangle}}{\langle k{\ell}\rangle}.
\end{eqnarray*}
\indent
Suppose $\alpha <1$. Then the threshold for any targeted immunization satisfies
\begin{eqnarray}
\lambda_{\rm c} = \frac{\langle \ell \rangle}
{\langle k{\ell}\rangle - \alpha {\langle k{\ell}\rangle}_T} \le \hat{\lambda}_{\rm c} 
= \frac{1}{1-\alpha}\lambda_{\rm c}^{\circ},
\end{eqnarray}
since $\langle k{\ell}\rangle_T \le \langle k{\ell}\rangle$ always holds.  
Therefore, in case of $\alpha <1$, certainly any targeted immunization is not more than the
global immunization. \\
\indent
Suppose that in addition to $\alpha <1$ considered networks are scale-free as presented in Section 1.
Scale-free networks are prone to have very large $\langle k{\ell}\rangle$;
$\lambda_{\rm c}^{\circ} \approx 0$, under a high correlation between nodal 
indegrees and outdegrees [11]. In such a case inequality (18) tells us that any targeted immunization is not
effective: $\lambda_{\rm c} \approx 0$.\\
\indent
Next supposing $\alpha = 1$, let us consider a simple target set $T$ by defining
$\bar{T}=\{(k, \ell) | k, \ell \le M\}$ for some positive $M$. This yields
\[
{\langle k{\ell}\rangle}_{\bar{T}}= \sum_{k, \ell \le M}k \ell p(k, \ell) < M^2,
\]
and hence, even in scale-free networks, we are able to conclude
\[
\lambda_{\rm c} = \frac{{\langle \ell \rangle}}
{{\langle k{\ell}\rangle} - {\langle k{\ell}\rangle}_T}
= \frac{{\langle \ell \rangle}}
{{\langle k{\ell}\rangle}_{\bar{T}}} >  \frac{{\langle \ell \rangle}}{M^2} >
\lambda_{\rm c}^{\circ} \approx 0,
\]
which implies the effectiveness of the targeted immunization only under the condition $\alpha = 1$.
The choice of a desirable target set $T$ 
depends on the probability distribution $p(k, \ell)$, so this estimate may be improved further.
\\
\indent
Finally, as for undirected networks, there is no distinction between indegrees and outdegrees.
So, setting $k = \ell$ in (15), the threshold of an epidemic outbreak can be written in our notation as
\[
\lambda_{\rm c} = \frac{\langle k \rangle}{{\langle k^2 \rangle}-\alpha{\langle k^2 \rangle}_T},
\]
for an immunization rate $\alpha$ and a target set $T$ that is characterized by the degree $k$.
Using this, we see that similar results of this section hold for undirected networks. Note that
in [12] the target set was assumed to be $T =\{k | k > K\}$ for a large number $K$, namely
a set of hub nodes. In this case the targeted immunization is not effective unless it confers
100 percent immunity on all hub nodes in $T$.
\\
\begin{center}
{\bf\large References}
\end{center}
\begin{itemize}
\item[{[1]}]  M. Abramowitz and I. A. Stegun (Eds.), {\it Handbook of Mathematical Functions}, 
Dover Pub., New York, 1972.
\item[{[2]}]  S. Boccaletti, V. Latora, Y. Moreno,  M. Chavez and D.-U. Hwang, 
Complex networks: Structure and dynamics,
{\it Physics Reports} {\bf 424}, 175--308, 2006.   
\item[{[3]}]  S. N. Dorogovtsev and J. F. F. Mendes,
{\it Evolution of Networks: From Biological Nets to the Internet and WWW}, 
Oxford Univ. Press, Oxford, 2003.
\item[{[4]}] X. Fu, M. Small, D. Walker and H. Zhang,
Epidemic dynamics on scale-free networks with piecewise linear infectivity and immunization,
{\it Physical Review} E {\bf 77}, 036113, 2008. 
\item[{[5]}] J. Giesecke, {\it Modern Infectious Disease Epidemiolgy}, E. Arnold Pub., London, 2002.
\item[{[6]}] Y. Moreno, R. Pastor-Satorras and A. Vespignani, 
Epidemic outbreaks in complex heterogeneous networks, {\it European Physical Journal B} 
{\bf 26}, 521-529, 2002.
\item[{[7]}] J. D. Murray, {\it Mathematical Biology}, Springer Verlag, New York, 2002.
\item[{[8]}]  M. E. J. Newman, The structure and function of complex networks,
{\it SIAM Review} {\bf 45} 167--256, 2003.     
\item[{[9]}]  R. Pastor-Satorras and A. Vespignani,
Epidemic spreading in scale-free networks, {\it Physical Review Letters} {\bf 86}, 3200--3203, 2001.
\item[{[10]}]  R. Pastor-Satorras and A. Vespignani, Immunization of complex networks,
{\it Phyical Review} E {\bf 65}, 036104, 2002.
\item[{[11]}]  S. Tanimoto, Epidemic thresholds in directed complex networks, arXiv:1103.1680, 2011.
\item[{[12]}]  Y. B. Wang, G. X. Xiao, J. Hua, T. H. Cheng and L. S. Wang,
Imperfect targeted immunization in scale-free networks, {\it Physica} A {\bf 388}, 2535--2546, 2009.
\end{itemize}
\end{multicols}
\end{document}